\documentclass[apjl]{emulateapj}

\usepackage{amsmath}

\newcommand{\hubble}{\emph{Hubble}}

\newcommand{\etal}{et al.}

\newcommand{\lya}{Lyman~$\alpha$}

\slugcomment{}

\shorttitle{} \shortauthors{}

\begin{document}

\title{Exoplanet HD\,209458\lowercase{b} 
(Osiris\footnote{B\lowercase{ecause the escape of hydrogen atoms is strengthened in this 
paper, we renew our proposal to use the nickname
"}O\lowercase{siris" for the planet} HD\,209458\lowercase{b who loses mass 
like the }E\lowercase{gyptian god.}}): Evaporation strengthened}

\author{A.~Vidal-Madjar$^{\rm 1,2}$, A.~Lecavelier des Etangs$^{\rm 1,2}$, J.-M.~D\'esert$^{\rm 1,2}$ \\
G.~E.~Ballester$^{\rm 3}$, R.~Ferlet$^{\rm 1,2}$, G.~H\'ebrard$^{\rm 1,2}$, and M.~Mayor$^{\rm 4}$}
\affil{$^{\rm 1}$CNRS, UMR 7095, 
   Institut d'Astrophysique de Paris, 
   98$^{\rm bis}$ boulevard Arago, F-75014
   Paris, France}
\affil{$^{\rm 2}$UPMC Univ. Paris 6, UMR 7095, 
   Institut d'Astrophysique de Paris, 
   Paris, France}
\affil{$^{\rm 3}$Lunar and Planetary Laboratory, University of Arizona, 1541 E. University Boulevard, 
Tucson, Arizona 85721-0063, USA}
\affil{$^{\rm 4}$Observatoire de Gen\`eve, CH-1290 Sauverny, Switzerland}

\begin{abstract}

Following re-analysis of \hubble\ \emph{Space Telescope} observations
of primary transits of the extrasolar planet HD\,209458b at \lya ,
Ben-Jaffel (2007, BJ007) claims that no sign of evaporation is
observed.  Here we show that, in fact, this new analysis is consistent
with the one of Vidal-Madjar et al. (2003, VM003) and supports the
detection of evaporation. 
The apparent disagreement is mainly due to the
disparate wavelength ranges that are used 
to derive the transit absorption depth. 
VM003 derives a $(15\pm4)$\%\ absorption depth during
transit over the core of the stellar \lya\ line (from $-130$~km/s
to $+100$~km/s), and this result agrees with the $(8.9\pm2.1)$\%\ absorption
depth reported by BJ007 from a slightly expanded dataset but over a larger
wavelength range ($\pm$\,200\,km/s). These
measurements agree also with the $(5\pm2)$\%\ absorption reported by
Vidal-Madjar et al.~(2004) over the whole \lya\ line from independent,
lower-resolution data. We show that stellar \lya\ variability is
unlikely to significantly affect those detections.
The H\,{\sc i} atoms must necessarily have velocities above the escape
velocities and/or be outside the Roche lobe, 
given the lobe shape and orientation. Absorption by H\,{\sc i}
in HD\,209458b's atmosphere has thus been detected with different datasets, 
and now with independent analyses. 
All these results strengthen the concept of evaporating hot-Jupiters, 
as well as the modelization of this phenomenon.

\end{abstract}


\section{INTRODUCTION}

\label{sec:intro}

Few detections of extrasolar planets' atmospheric species are
reported so far, but they have been recognized as important steps in our
understanding of these objects. Of particular
interest are the direct detections with the {\sl Hubble Space Telescope} (HST)
during primary transits of
NaI by Charbonneau \etal\ (2002) as well as HI, OI and CII by
Vidal-Madjar \etal\ (2003, 2004) (hereafter VM003 and VM004) 
and Ballester \etal\ (2007).

The recent paper by Ben Jaffel (2007) (hereafter BJ007) however
casts some doubt on many aspects of the HI detection in the upper
atmosphere of HD\,209458b and on the implication that the planet is
evaporating due to the large energy input from its nearby host
star (e.g., Lecavelier des Etangs 2007). 
Consequently, a related large number of theoretical studies 
would have to be revised.

In the present rebuttal paper, we discuss the BJ007 
arguments and show where they are misguiding.

\section{Where is the difference ?}
\label{sec:diff}

BJ007 completed a new data analysis based on sampling the same
observations as VM003 in a different temporal manner, and adding 
data from two HST orbits 
from an archival program completed at another epoch (two orbits added
to the nine HST orbits of the VM003 observations). From the resulting new
\lya\ transit light curve, the transit absorption depth is evaluated by
integrating the \lya\ flux in two ranges~: ``blue'' (1214.83-1215.36\AA) 
and ``red'' (1215.89-1216.43\AA). 
This slightly increases the wavelength domain excluded from the
analysis because of geocoronal contamination, from
1215.5-1215.8\AA\ in VM003 to 1215.36-1215.89\AA\ in BJ007. This
has, however, no significant consequences on the evaluation as shown in the
Figure~4 of VM003.

BJ007 evaluates the transit depth by fitting a light curve 
over the observations sampled by 300s sub-exposures as
a function of the planet orbital phase. The resampled observations
show some level of variability mainly due to the variation of the
stellar \lya\ flux. The fitting procedure ``smooths'' these
variations as if all observations were made with an average
stellar \lya\ flux. This gives an average HI absorption during the
transit of ($8.9\pm 2.1)\%$ by considering a wavelength range from
1214.83 to 1216.43~\AA\ which corresponds to $\sim \pm 200$~km/s
in velocity space.

The data analysis of BJ007 is not put into question here. The major
differences between BJ007 and VM003 are on the data
interpretation. Both analyses provide a \lya\ line flux as a
function of time and wavelength, which includes possible stellar
variations and planetary transit signature. The differences lie i) in the
wavelength ranges used to convert the
spectra, as a function of time, into a single absorption depth
measurement and ii) in the reference flux used to correct 
for the intrinsic stellar flux variations.

\begin{figure}[tbh]
\resizebox{\hsize}{!}{\includegraphics[angle=0]{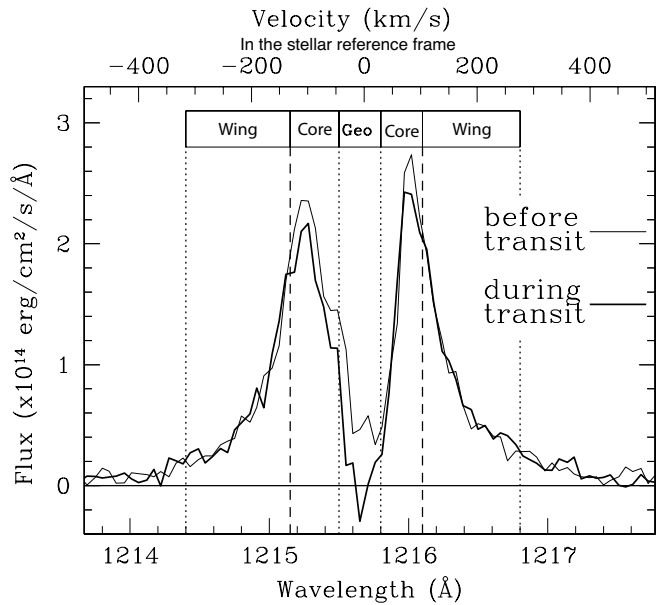}}
\caption{The observed HD\,209458 \lya\ profiles as observed by
Vidal-Madjar \etal\ (2003) before and during the planetary transit. 
The BJ007 re-analysis of nearly the same data set
produced a similar \lya\ line profile. 
The two vertical dashed lines define
the limits $\lambda_1$ and $\lambda_2$ of the line core 
where HI planetary absorption takes place. In VM003 as well as in BJ007, the
central part of the line (noted ``Geo'') possibly perturbed by the
Earth geocoronal emission is omitted from the analysis. 
The line wings 
are used by VM003 as a flux reference to correct 
for the stellar \lya\ intrinsic variations.} \label{fig:LyA}
\end{figure}

The \lya\ emission of the star can significantly vary
from epoch to epoch.
Because there is no detectable transit absorption signature in the \lya\ 
line wings (the nominal 1.5\% obscuration by the planetary
disk is below the data S/N), VM003 calculated a \emph{relative} absorption
depth using wings of the \lya\ line as flux reference. This method
aims at correcting for any intrinsic and unknown changes in the
\lya\ stellar flux (see \S~\ref{stellar variations}). 
As illustrated in
Fig.~\ref{fig:LyA}, VM003 defined two spectral domains~: the line
core from $\lambda_1$ to $\lambda_2$ 
(called ``In'' in VM003 and ``Core'' in Fig.~\ref{fig:LyA}; 
excluding the part of the spectrum contaminated by the geocoronal
\lya\ emission), and the remaining wavelength domain in the wings
of the line used as a flux reference from 1214.4\AA\ 
to $\lambda_1$ and from $\lambda_2$ to 1216.8\AA\ 
(called ``Out'' in VM003 and ``Wing'' in Fig.~\ref{fig:LyA}). The
best domain defined by $\lambda_1$ and $\lambda_2$ is found by
varying the positions of these two wavelengths until the strongest
absorption signal is identified. VM003 found that the strongest
absorption relative to the line wings takes place in the line core
between $\lambda_1$=1215.15 and $\lambda_2$=1216.10\AA\ 
(excluding the central geocoronal region).

\begin{table*}
\caption{Evaluated \lya\ absorption over various spectral
domains \label{tab:fit}}
\begin{center}
\begin{tabular}{lccc}
\hline
\hline
\noalign{\smallskip}
            & Line core & Intermediate & Whole line \\
\noalign{\smallskip}
Spectral domain & 1215.15-1216.10\AA & 1214.83-1216.43\AA & 1210-1220\AA \\
& VM003 limits & BJ007 limits &   \\
\noalign{\smallskip}
\hline
\noalign{\smallskip}
Published absorption (\%) & $15\pm 4$ &  $8.9\pm 2.1$ &  $5\pm 2$ \\
(reference and data set)  &     (VM003)    &    (BJ007)&    (VM004)  \\
\noalign{\smallskip}
\hline
\noalign{\smallskip}
Absorptions using VM003 dataset$^{\rm a}$ & 
$15.1\% \pm 4\%$ &  $7.3\%\pm 2.0\%$ & $5.7\%\pm 1.9\%$ \\
\noalign{\smallskip}
\hline
\hline
\noalign{\smallskip}
\multicolumn{3}{l}{$^{\rm a}$ Absorption evaluated using VM003 dataset }\\
\multicolumn{3}{l}{and following the VM003 method with wings as flux reference.} \\
\end{tabular}
\\
\end{center}
\end{table*}

Because the HI absorption takes place only in the central part of
the \lya\ line, the absorption depth measurement must decrease
for increasing wavelength range, including wings of the line 
where there is no absorption. 
This has already been found by VM004 
with an independent data set obtained 
with STIS in the G140L low resolution mode. In the low resolution data, 
the \lya\ stellar emission line is not spectrally resolved and only the
total \lya\ flux can be evaluated. 
VM004 obtained a $(5\pm2)\%$ transit absorption depth over the whole line,
in agreement with the estimate obtained from the 
VM003 dataset (Table~\ref{tab:fit}).

To compare the VM003 result to that of BJ007, we can also
calculate the transit absorption depth over the spectral domain as defined by BJ007 
using the spectra of VM003. 
Keeping the same approach as in VM003 to 
account for stellar \lya\ variations, we evaluate the planetary 
absorption during transit using the line wings as flux reference. 
We find a mid-transit absorption of (7.3$\pm$2.0)\% over 
the same spectral region used by BJ007, which is 
in agreement with the
BJ007 result of 8.9\% considering that the data set is the same, 
except for the addition of two HST orbits to the nine used by VM003.
This result shows that, considering the same wavelength range, similar
transit absorption depths are found in the VM003 and BJ007 spectra
(Table~\ref{tab:fit}). 

The uncertainty and noise appear to be lower in BJ007 than
in VM003. The difference is explained by the larger wavelength 
range used by BJ007 to estimate the \lya\ flux and 
by the uncertainty introduced by the correction of the \lya\ variations 
applied by VM003 (see \S~\ref{stellar variations}).

In short, the results given in VM003 and BJ007 for \lya\ absorption depths 
can be reconciled. The apparent difference is basically due to
the width of the spectral domain over which the absorption is
computed, acknowledging that the HI absorption does not
cover the whole extent of the stellar \lya\ line.
When using a larger wavelength domain, the absorption signal is
diluted and the absorption depth measurement is lower, 
as found in BJ007 compared to VM003, and in VM004 compared to 
BJ007. 

\section{High velocity blue-shifted absorptions}

Another argument, made by BJ007 against the evaporation scenario,
is that the blue-shifted absorption (produced by
hydrogen atoms at speeds up to $-130$km/s identified by VM003) 
is not confirmed in the BJ007 analysis. 
Velocities of $-130$km/s are above
the escape velocity of about $\sim$42\,km/s (at 1~$R_p$). 
If observed, high velocity atoms must be escaping
the planet.

Following the same approach as above using the VM003 dataset, 
we evaluate the absorption seen during transit within the ``blue'' and the
``red'' sides of the wavelength domain defined by BJ007.
In this domain, we find (9.8$\pm$1.8)\%\ and (5.2$\pm$1.0)\%\ absorption 
in the ``blue'' and the ``red'' sides, 
respectively. Therefore, the large spectral domain defined 
by BJ007 (including velocities up to $\pm 200$\,km/s)
shows a significant absorption in both blue and red sides 
with, as found in VM003, an absorption stronger in the
blue than in the red.

Finally, although very little work has been published to explain
these high velocities, they can be produced by radiation pressure 
from the intense Lyman-$\alpha$ flux of the nearby star 
(see the velocity diagram and the cometary shape of the escaping atoms
in Fig.~3 of Vidal-Madjar \& Lecavelier des Etangs (2004)).
In which case, a difference between blue and red absorptions is expected.

\section{The Lyman-$\alpha$ stellar variations}
\label{stellar variations}

In previous works, corrections for stellar variations was
done either using the wings of
the line (VM003) or the observations before and after the
transit (VM004), while BJ007 correction was done by
averaging the flux variations by
fitting a standard transit curve over the phase-folded data points
(see Fig.~2 in BJ007).

Contrary to the claim made by BJ007, the \lya\ variations of HD\,209458
are not necessarily relatively large but normal for a quiet solar type
star (Vidal-Madjar, 1975). In fact, when data are phase-folded, 
the apparent variations can be artificially enhanced due to the superimposition 
of observations made at different dates. Moreover, real stellar
fluctuations are combined with significant noise which is expected 
to be large because of the limited number of photons
in the short temporal bins.

\begin{figure}
\resizebox{7cm}{!}{\includegraphics[angle=0]{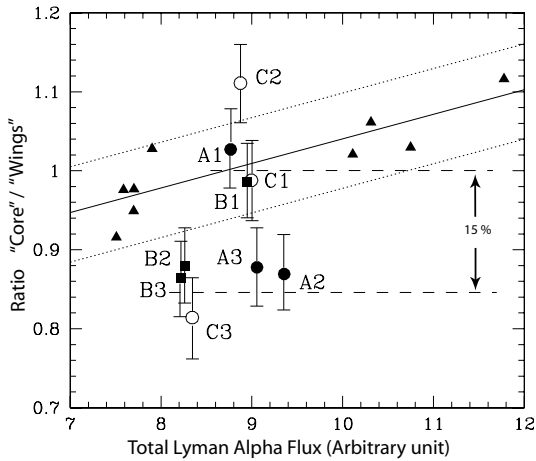}}
\caption{Plot of the Solar and HD\,209458 ``Core''/``Wing'' ratios as a
function of the total \lya\ flux (``Core''+``Wing'').
``Core'' and ``Wing'' domains are defined in Fig.~\ref{fig:LyA}.
Triangles are for Solar values. Circle and square symbols
are for HD\,209458
and correspond to the three HST visits A, B, and C, each visit
including three successive HST orbits, namely 1, 2 and 3 
(see VM003 for details). The linear regression to the Solar
values is shown along with the upper and lower limits (dotted lines). 
The HD\,209458 values show basically two groups. 
In a first group, out-of-transit measurements A1,
B1 and C1 are within the limits
of the Solar fluctuations. The second
group consists of in-transit measurements (where A2, B3 and C3 are the closest mid-transit) 
that are well below the Solar fluctuations. 
The comparison between the dispersion within a group and 
the amplitude of the transit
signature between the two groups, the absence of correlation between the HD\,209458b 
``Core''/``Wing'' ratios and the total \lya\ flux, both 
invalidate the suggestion that 
HD\,209458 \lya\ activity corrupts the transit signature.} \label{fig:Sun_var}
\end{figure}

The Sun can be used as a proxy to study the possible variations of the flux and shape
of the HD\,209458 \lya\ line. HD\,209458 is a Solar type star (G0V) 
whose similarity with the Sun is demonstrated by the
observation of the chromospheric CaII line profiles, which are relatively
quiet for both stars. From observations obtained with SOHO (Solar and
Heliocentric Observatory, Lemaire \etal , 2002), we show that
the core-to-wing ratio of the Solar \lya\ line
varies by about $\pm8\%$, while it varies by about $\pm6\%$ for a
given total \lya\ flux (Fig.~\ref{fig:Sun_var}). 
HD\,209458 is expected to have the same behavior and the 15\%\ 
transit measurements is unlikely to be due to stellar variations.
This is strengthened by the HD\,209458 \lya\ measurements
which present two different groups of core-to-wing 
flux ratios (Fig.~\ref{fig:Sun_var}) :
the ratios measured before the transit  
and the ratios during the planetary transit.
Although these measurements are obtained at various epochs,
the dispersion within a given group is small and within 
the error bars, while the difference between the
measurements taken during the transit compared to the 
reference measurements taken before the transit is significant
and larger than variations observed in the Sun. 
This behavior is unlikely to be coincidental and corresponds to
the expected signal for a cloud of HI atoms in
the environment of HD\,209458b.

Moreover, the dispersion of these individual measurements
shows no correlation with the total \lya\ flux. 
This shows that the stellar variations are unlikely 
to be responsible for the observed \lya\ core-to-wings 
variations. The core-to wing variations are thus
more likely related to the planetary transit.

Finally, it is extremely unlikely that stellar variations 
can mimic a transit light curve as seen
in Fig.~2 of BJ007, when measurements are phase-folded 
with the planetary orbital ephemerides. 

In summary, the stellar variations 
are taken into account by VM003 by comparing 
the variations in the core to the variations in the wings of the line, 
and by BJ007 by averaging a smooth transit light curve over a fluctuating \lya\
stellar flux. The similarity of the resulting absorption depths, 
using the BJ007 and VM003 dataset, 
when evaluated over the
same spectral domain (Table~1) shows that the
\lya\ stellar variability does not corrupt the transit evaluation
whatever the approach used for the data analysis.

\section{The size of the absorbing cloud and the Roche lobe}

BJ007 compares the $8.9\%$ absorption depth derived in his work 
to the absorption
caused by an occulting disk with a size of $\sim4.08$ R$_P$
supposed to be the size of the Roche lobe as calculated using
equations found in Gu et al.\ (2003). In the case of HD\,209458b,
a disk with a radius of $\sim4.08$ R$_P$ corresponds to an
absorption depth of about 25$\%$ during transit. BJ007 thus
concludes that the observed hydrogen atoms are inside the Roche
lobe and cannot escape the planet.

First, as discussed above, the BJ007 evaluation of $\sim 9\%$ is 
only a fraction of the full 
HI absorption depth in the line core. 
Second, the formula of Gu et al. (2003) for computing the Roche
lobe corresponds to the Lagrangian point L1 between the star and
the planet, i.e. the most distant Roche limit position relative to
the planet (see Eq.~B.8 and 
discussion in Lecavelier des Etangs, 2007). 
However, the Roche lobe around the planet is not spherical but elongated toward
the star 
(see e.g. Lecavelier des Etangs \etal\, 2004). In a transit configuration,
the observed limit of the Roche lobe is in a direction perpendicular
to the star-planet direction. In this perpendicular direction, 
the Roche lobe extension is about 2/3 of the extension toward
the L1 point. Therefore, it is more appropriate to use 
an average distance to the Roche lobe, which was given 
by VM003 to be $\sim2.7$ R$_P$ (Paczynski, 1971). 
A filled Roche lobe corresponds to about 12\%\ absorption.
This value is comparable to the HI observation of $\ga 10\%$
absorption depth in the line core. We can conclude
that HI atoms reach the Roche lobe or beyond, in agreement
with the models of atmospheric escape.

In addition, although observation of hydrogen atoms outside the 
Roche lobe is a direct evidence for escape, it is not a necessary 
condition. Even filling-up half a Roche lobe would imply 
escape rates large enough to significantly affect atmospheric 
structure (Lecavelier des Etangs et al.\ 2004). 
Thus, whether the hydrogen cloud actually fills up the Roche lobe or not, 
the large extension of the upper atmosphere (and possibly 
a cometary shape due to radiation pressure) shows that the 
atmosphere is indeed escaping.

\section{Consequence: Evaporation is confirmed}
\label{sec:consequences}

We have demonstrated that the BJ007 analysis is in agreement with
HI escape from the HD\,209458b atmosphere. Even if one considers
that the absorption value given by BJ007 is correct, then either: 
i) this $\sim$9\%\ absorption depth is taking place over the whole
spectral range of $\pm200$\,km/s, or ii) the $\sim$9\%\ absorption 
is the result of an unresolved larger absorption 
within a narrower wavelength range, produced by atoms below the escape
velocity of about $42$km/s. In the first case, HI atoms are detected 
to move at velocities much larger than the escape velocity.
In the second case, the absorption takes place over a narrow
spectral range, $\sim$5 times narrower than the range considered 
in BJ007. In this narrow wavelength range, the absorption must be $\gg$12\%\ 
and HI atoms must be present beyond the Roche lobe.
In both alternatives, the result of the data analysis described in BJ007
shows that HI must be escaping the planet atmosphere.

\section{Conclusion}
\label{sec:conclusion}

BJ007 called into question the VM003 discovery of 
the atmospheric escape from the HD\,209458b extrasolar planet. 
BJ007 gives two main arguments~: i)
the absorption depth is smaller than previously estimated 
and the Roche limit is not reached, and ii) the data analysis 
is corrupted by intrinsic stellar \lya\ variations.
The first argument is not correct because the absorption 
is not taking place over
the whole \lya\ line. The absorption
depth measurements depend on the considered wavelength range. 
In addition, the Roche lobe shape and orientation must be taken
into account when comparing the size corresponding to these absorption 
depths with the size of the Roche lobe.
The second objection is rejected by recalling that
VM003 corrected for the stellar variations using the overall 
stability of the line shape and the line wings as flux reference. 

Finally, despite the affirmation in the introduction
of BJ007, works to transpose these observations into estimates
of escape rate have been made early in VM003 and 
recently in Schneiter et al.\ (2007). 
As stated by our anonymous referee, 
these ``observational'' estimates agree with the most recent 
and sophisticated models which all find rather good agreement 
with mass loss rates of $\sim 3-7\times 10^{10}$~g\,s$^{-1}$. 
Therefore, at the orbital distance of HD\,209458b, 
atmospheric escape does not strongly affect the 
evolution of the planets, in accord with the evolutionary studies.

Last, but not least, Ben Jaffel (2007) performed an independent, 
thorough and careful analysis of the best available data set 
for measuring the \lya\ transit absorption depth. Despite a 
misinterpretation of the resulting spectra, this work
confirms that the signal detected in VM003 is indeed present
and is not related to the data reduction process. 
Therefore, in the atmosphere of HD\,209458b, 
HI appears to be the only species to have been 
detected with two independent datasets, 
and now with independent data analyses.
The BJ007 data analysis strengthens the detection of the extended 
escaping atmosphere of HD\,209458b, in agreement with numerous 
theoretical studies of this phenomenon (e.g., Lammer et al. 2003; 
Lecavelier des Etangs et al. 2004; Yelle 2004, 2006; 
Baraffe et al. 2005; Tian et al. 2005; Garcia-Munoz 2006).

\begin{acknowledgements}
We thank the referee for constructive and strengthening remarks. 
This work is based on observations
with the NASA/ESA Hubble Space Telescope, obtained
at the Space Telescope Science Institute, which is operated by
AURA, Inc.
\end{acknowledgements}

\end{document}